# MAGNETIC SHIELDING FOR INTERPLANETARY TRAVEL

**M.W. SAILER[1] AND H.M. DOSS[2]**

[1]*Matthew Sailer 1602 Governors Dr, APT 1627, Pensacola, Fl 32514*
[2]*Point Loma Nazarene University, Physics & Engineering Department, 3900 Lomaland Dr., San Diego, CA 92106*
**Email:** mwsailer12@gmail.com[1] and hdoss@pointloma.edu[2]

A proposed design for radiation shielding in interplanetary travel is presented with primary shielding created by a super conducting split toroid magnetic field and unconfined magnetic fields created by two deployable superconducting loops. The split toroid's shielding effectiveness is analyzed with calculations of mass, field, particle path, and synchrotron radiation provided. The creation of plasma regions is also taken into consideration. The design has a wire mass of $8.08 \times 10^4$ kg, which creates a 0.47 T shielding field, and a field less than $2.17 \times 10^{-4}$ T in the crew area. Calculations show the shielding field deflects concerning particles of solar wind, solar flares, and high energy high atomic number particles found in galactic cosmic radiation. This design might also be used to generate power and thrust from the radiation to create a potentially self-sufficient mobile station.

## 1. Introduction

Interplanetary travel requires radiation shielding to protect astronauts and electronics. The radiation of most concern is that from the solar wind and solar flares, which produce solar energetic particles (SEP) and the radiation produced in galactic events such as supernovae known as galactic cosmic radiation (GCR). SEP radiation consists mainly of protons and electrons. GCR consists of ionized H, He, as well as high atomic number energetic particles (HZE) such as $Fe^{+26}$, $O^{+8}$, and $C^{+6}$ [1]. To shield these particles both their maximum energy and the energy of maximum flux need to be taken into consideration [2], as well as secondary radiation that might be produced during shielding.

Passive methods that have been considered include mass shielding [3], and active methods considered include creating plasmas [4], electric fields [5], and magnetic fields [6-11] surrounding the system. No one of these systems seems to be enough to sufficiently reduce radiation exposure. Ongoing research in new passive materials such as BNNT [3] is very promising as is the application of superconducting materials and new designs for fields [12-13].

This work proposes a combination of both passive mass shielding and active magnetic shielding with consideration of trapped plasma as a design. The focus of this paper is on the magnetic field created by the design and calculations of particle trajectories through the field with consideration of the secondary radiation the particles produce.

## 2. Proposed Design

A shielding method consisting of a split toroid wire configuration, two deployable wire loops, and passive shielding using Boron Nitride Nanotubes (BNNT) is proposed. The crew area is a cylindrical shape with a 10 m radius, and a 10 m height as seen in Fig. 1. Its surface is constructed from BNNT to minimize any radiation that might make it through the fields and any produced secondary radiation from reaching the crew area. The thickness of this BNNT wall can be adjusted based on shielding effectiveness of the material and weight, but it is shown as 1 m in Fig. 1.

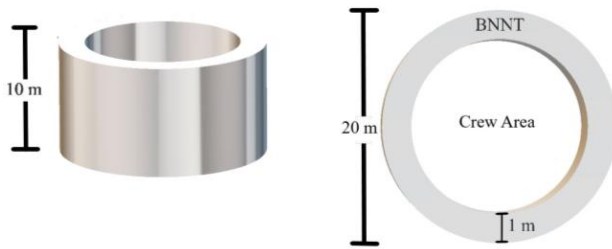

Figure 1: Crew Area Dimensions

Superconducting wire is necessary to sustain a $B$-field during interplanetary travel. Ti-MgB$_2$ Superconducting tape has a high current to mass ratio as Musenich et al. and Buzea and Yamashita demonstrate [12-13]. Tables 1-2 show the limits of the Ti-MgB$_2$ Superconducting tape according to both articles.

| $B$-field | 0.25 T | 0.5 T | 1.0 T |
|---|---|---|---|
| Max Current | 885 A | 788 A | 594 A |
| Temperature | 16 K | 16 K | 16 K |

Table 1: Ti-MgB$_2$ Musenich et al. data 2015 calculated from Ti-MgB$_2$ tape – improved process

| $B$-field | 0.25 T | 0.5 T | 1.0 T |
|---|---|---|---|
| Max Current | 8600 A | 7200 A | 5800 A |
| Temperature | 25 K | 25 K | 25 K |

Table 2: MgB$_2$ Buzea and Yamashita data 2001

Tables 1-2 have data differing by about one order of magnitude. This difference comes from the titanium sheath used by Musenich et al., which allows the avoidance of helium cryogenics and maintains stability of the tape [12]. Table 2 shows the potential MgB$_2$ superconductors have as a wire for $B$-fields as radiation shields in space. Until more work is done, this paper will assume the values from the data of Musenich et al. as the limits for MgB$_2$ superconductors. The Ti-MgB$_2$ Superconducting tape will be referred to as a wire for the rest of this paper for simplicity.

A split toroid made from Ti-MgB$_2$ superconducting wire creates a roughly uniform, 5 m thick, $B$-field of 0.47 T that surrounds the crew area. The inner loops have a radius of 10 m while the outer loops have a 15 m radius as shown in Fig. 2.

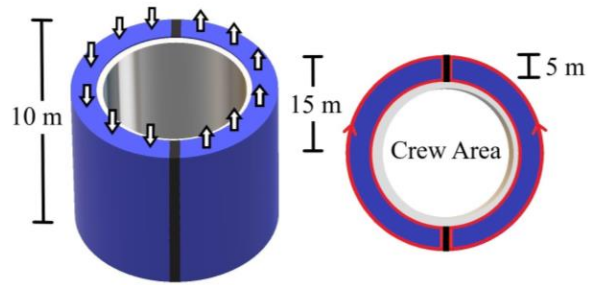

Figure 2: Ti-MgB$_2$ Split Toroid $B$-field surrounding the crew area

The superconducting wires of the split toroid carry a current of 800 A at a temperature of 16 K, with a wire density of 600 wires per meter extending vertically 10 meters, surrounding the crew area. A wire density of this magnitude will have significant wire repulsive forces associated with the toroidal loops and will need to be optimized in a final design by adjusting the distance between each wire. Toroidal caps deflect particles at the top and bottom of the split toroid with the same current and wire density making the split toroid a nearly confined $B$-field, however, their shape is adjustable as shown in Fig. 3.

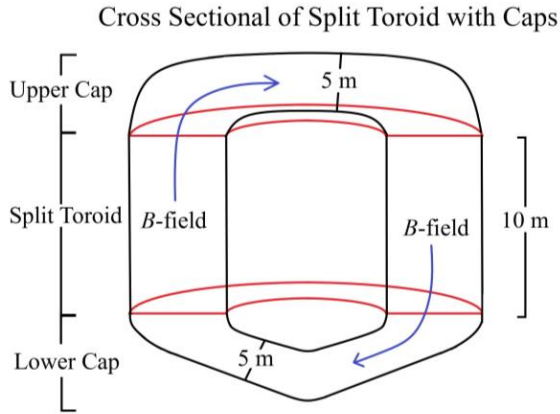

Figure. 3 Split Toroid with Adjustable Caps

A deployable wire loop of radius 100 m, current 800 A, and $n_o$ = 90 superconducting loops of wires carrying a total combined current of $n_o I$ = 72,000 A surrounds the split toroid and produces a *B*-field with similar magnitude to that of Earth. As demonstrated by the SR2S project, [10] a plasma of trapped particles forms surrounding the deployable loops similar to that of the Van Allen belts, Fig. 4.

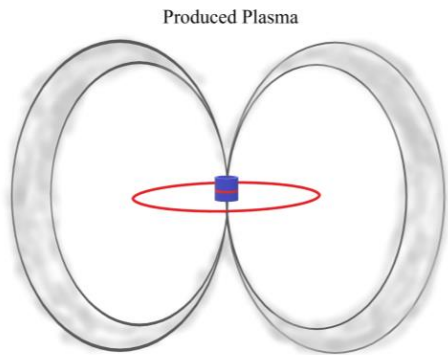

Figure 4: Induced Plasma from *B*-field

A smaller wire loop of 15 m radius, current 800 A, and $n_i$ = 20 superconducting wires loops carrying a total combined current of $n_i I$ = 10,800 A will cancel out the *B*-field in the crew area as seen in Fig. 5.

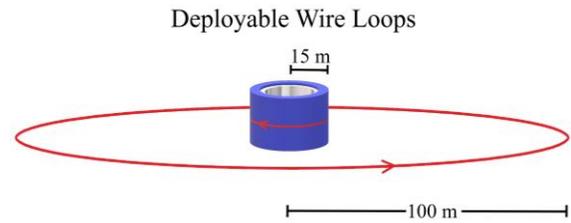

Figure 5: Two Deployable Wire Loops of Opposite Current shown in red

The adjustable toroidal caps have potential of producing thrust from redirected radiation spiraling down at the top and bottom of the crew area as shown in Fig. 6. The direction of thrust can be adjusted by manipulating the shape of the caps to change the direction of deflection of the particles.

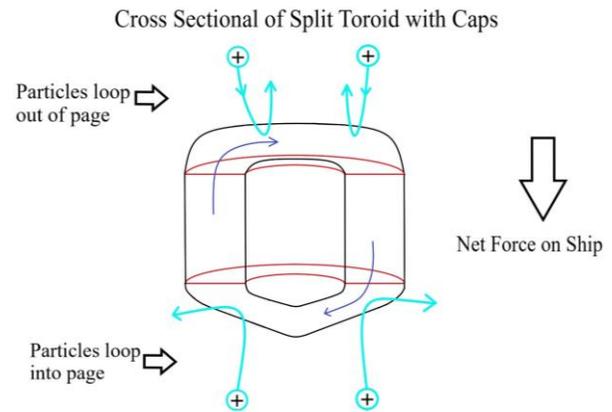

Figure 6: Particle Redirection at Caps

Produced synchrotron radiation may be harnessed for energy with the incorporation of photovoltaic cells [14].

While there are many parts to this design, this paper focuses on the shielding effectiveness of the *B*-field created by the split toroid and the two deployable wire loops.

## 3. Calculations

Before any simulations were run, highly energetic "test particles" were chosen as a goal for the shielding method to deflect. Based on Miller and Zeitlin's data, particles with twice the energy of the most probable GCR particles were chosen. These particles were $Fe^{+26}$, $C^{+6}$, $He^{+2}$, and $H^+$ with kinetic energies of 84 GeV, 18 GeV, 12 GeV, and 1.5 GeV respectively. [2].

A $B$-field simulation was created using MATLAB coding with the Biot-Savart law in Eq. 1.

$$\vec{B} = \frac{\mu_0 I}{4\pi} \int \frac{d\vec{l} \times \hat{r}}{|r|^2} \qquad (1)$$

The following information was input into the code: Matrix dimensions, split toroid dimensions and specifications, and deployable loop currents and dimensions. The code outputs a 3-dimensional vector field using matrices. The $B$-field vector was calculated, from each wire in the split toroid and the deployable loops, at each point in the matrix.

A second MATLAB code was used to send virtual particles into the $B$-field from various directions and observe their path. This code has inputs of particle mass, charge, initial kinetic energy, and the $B$-field matrix created by the toroid and deployable loops. The relativistic particle's path is plotted using time steps which calculate acceleration, new velocity, new position, produced synchrotron radiation, and change in kinetic energy before moving on to the next time step and continues in this fashion until the particle leaves the vector field. We used a time step of $0.1 \text{ m}/|v|$ where $|v|$ is the magnitude of the initial velocity of the time step which gives a reasonable estimate of the particle's path. The particle's path is calculated based on the value of the nearest $B$-field vector to provide an accurate trajectory as the particle travels through the field. The code outputs the total produced synchrotron radiation in electron volts, and visual plots of the particle's trajectory, the $B$-field, and the toroidal and deployable wires.

Calculations performed to update variables are given by Eq. 2 for acceleration $\vec{a}$, Eq. 3 for velocity $\vec{v}$, Eq. 4 for position $\vec{r}$, and Eq. 5 for the synchrotron radiation power, $P$. In what follows, $m$, represents the rest mass of the particle, $t$, the time that has passed, $c$, the speed of light in vacuum, $q$, the charge of the particle, and

$$\gamma = \left(1 - \frac{v^2}{c^2}\right)^{-1/2}$$

$$\vec{a} = \frac{q}{m\gamma}\left(\vec{v} \times \vec{B}\right) \qquad (2)$$

$$\vec{r} = \vec{r}_0 + \vec{v}t + \frac{1}{2}\vec{a}t^2 \qquad (3)$$

$$\vec{v} = \vec{v}_0 + \vec{a}t \qquad (4)$$

$$P = \frac{q^2 \gamma^4 c}{6\pi \epsilon_0 r^2} \qquad (5)$$

The total energy lost per particle due to synchrotron radiation was calculated by summing over the total time the power radiated multiplied by the time step.

To calculate the $B$-field, a $200 \times 200 \times 200$ matrix was created with each vector representing one square meter. A minimum toroidal radius of 10 m and a maximum radius of 15 m was input as the dimensions of the split toroid to create the toroid seen in Fig. 2. A current of 800 A was sent through the toroidal loops using 6000 layers of wire over a vertical distance of 10 m to give an estimate of the field produced. A deployable loop of 90 wires was added with a radius of 100 m. A second deployable loop of opposite current and a radius of 15 m was added to cancel out the $B$-field in the center from 20 wires creating the configuration seen in Fig. 5.

$Fe^{+26}$, $C^{+6}$, $He^{+2}$, $H^+$, and $e^-$ particles were sent into the *B*-field in groups of ten of each particle type from the bottom left corner at the matrix points [1 1 101] to [10 1 101]. The particle's velocity is directed towards the center of the crew area to maximize the amount of *B*-field traversed. Due to the unconfined toroidal B-field in the simulation, the particles are initially redirected to a tangential path along the toroid and some are pulled into the toroid as grazing particles. In a final design, this effect will not occur to this extent due to the toroid being a nearly confined field. Different kinetic energies were input until a maximum kinetic energy for a grazing particle was found that would not enter the crew area. The values of synchrotron radiation produced were also calculated.

The *B*-field was then multiplied by $\frac{1}{4}, \frac{1}{2}$, 2, and 4 to calculate its shielding effectiveness for a total of 5 different configurations. These values will be called *B*-field Multipliers in this paper. The corresponding mass was also calculated for each *B*-field Multiplier. The shielding effectiveness of each *B*-field Multiplier is calculated in the following section.

To calculate the total mass of the wires, the dimensions of the Ti-MgB$_2$ superconductors were used and a mass per length of 0.037 kg/m [12]. The total length of wire of the deployable loops and split toroid was added up to calculate a wire mass of $4.15\times10^4$ kg. The total mass, including an estimate of the toroidal caps, is calculated to be $8.08\times10^4$ kg by doubling the mass of the toroid wires.

## 4. Results

The resulting *B*-field from the deployable loops is similar in magnitude to Earth's geomagnetic field, but spread out over a smaller volume. This was calculated based on representing Earth's outer core with the larger deployable wire loop and choosing a current to produce the same *B*-field magnitude at a distance of 220 m from the center of the loops representing Earth's surface. The resulting field is shown in Figs. 7-9.

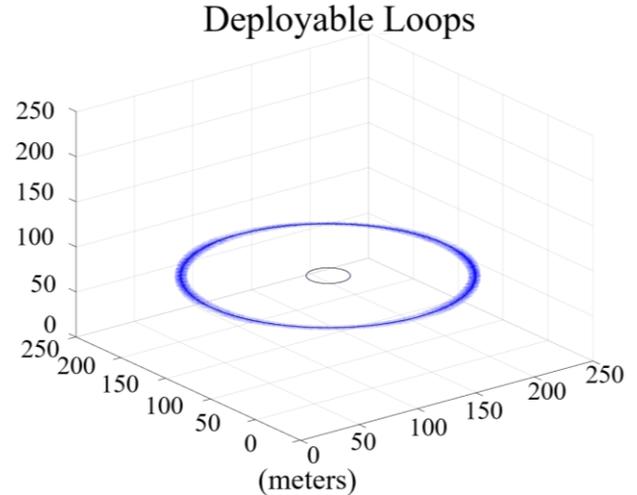

Figure 7: Deployable wire loops creating *B*-field of similar magnitude to Earth

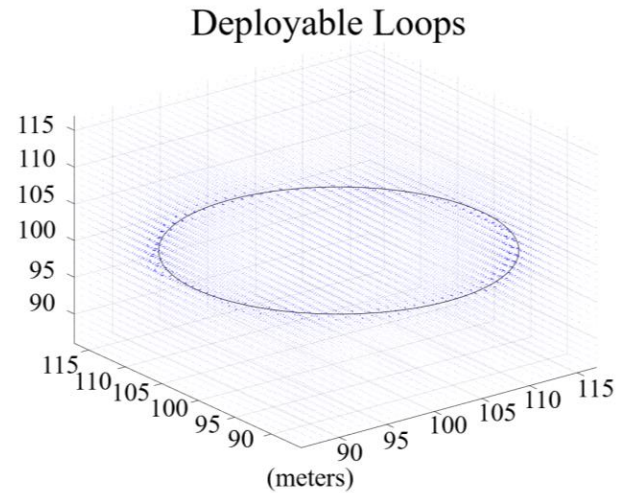

Figure 8: Center of *B*-field from deployable loops (side view)

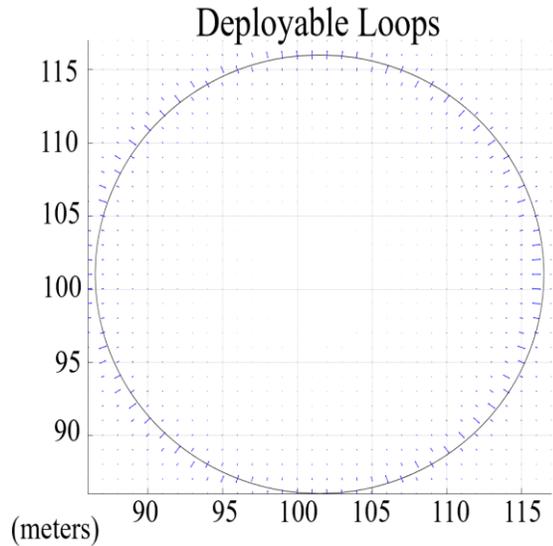

Figure 9: Center of *B*-field from deployable loops (top view)

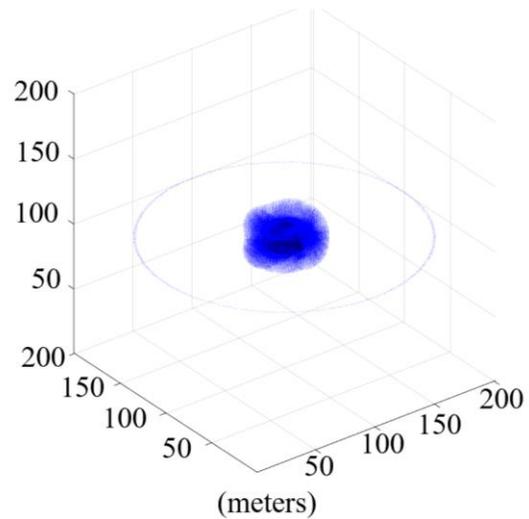

Figure 10: Split Toroid and Deployable Loop *B*-field (side view)

The maximum *B*-field produced by the deployable loops inside the crew area's 10 m radius is $2.17 \times 10^{-4}$ T with the center not exceeding $8.63 \times 10^{-5}$ T. These are acceptable values according to the International Standards of static *B*-field prolonged exposure from the World Health Organization [15].

The split toroid *B*-field was then calculated without the addition of caps and added to the deployable loop's *B*-field. Because the caps were not added to this calculation, the resulting *B*-field for the toroid is not fully confined. The additional field in the crew area, created from the split toroid, is neglected in the total *B*-field of the crew area since the final design will be nearly fully confined. This combined *B*-field can be seen in Fig. 10-12.

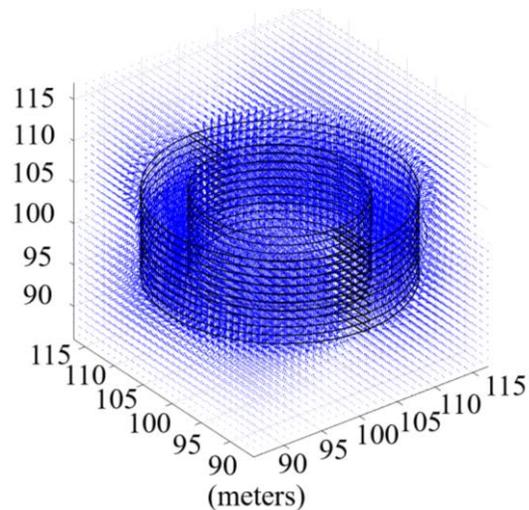

Figure 11: Split Toroid *B*-field (side view)

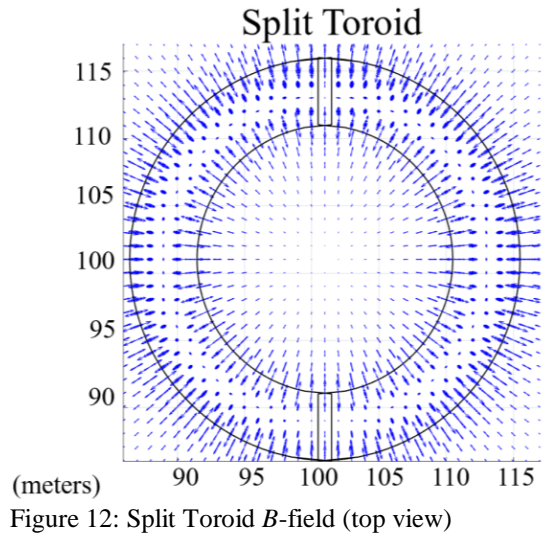

Figure 12: Split Toroid *B*-field (top view)

Particles in groups of ten were sent into this field from the bottom left corner from 1-10 m on the *x*-axis as previously described, and their paths were plotted in red as seen in Figs. 13-14. Note that MATLAB has a tendency to connect the final positions of each particle with a straight line to each other. This MATLAB artifact does not obscure the path of the particles. When only one particle is sent through, no extra lines appear to create a clearer picture of the particle's path as seen in Fig. 15.

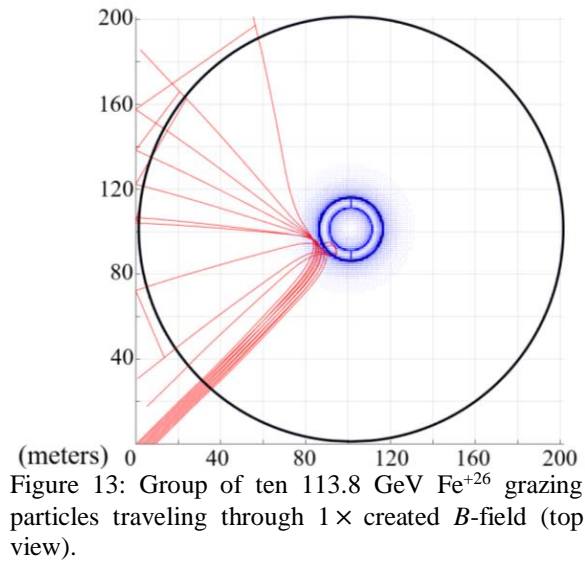

Figure 13: Group of ten 113.8 GeV $Fe^{+26}$ grazing particles traveling through 1× created *B*-field (top view).

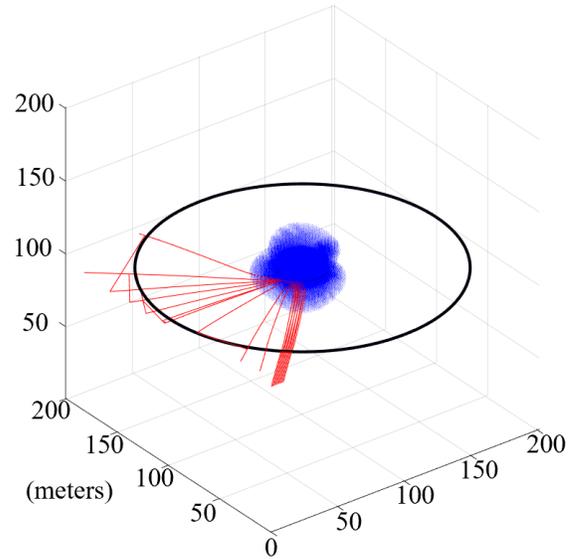

Figure 14: Group of ten 113.8 GeV $Fe^{+26}$ grazing particles traveling through 1× created *B*-field at (side view)

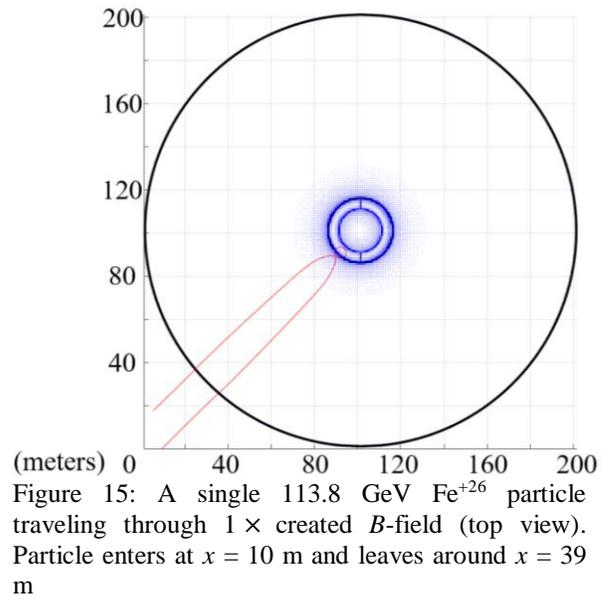

Figure 15: A single 113.8 GeV $Fe^{+26}$ particle traveling through 1× created *B*-field (top view). Particle enters at $x = 10$ m and leaves around $x = 39$ m

Particles grazing the outer deployable loop were simulated to demonstrate that they are also shielded as shown in Fig. 16.

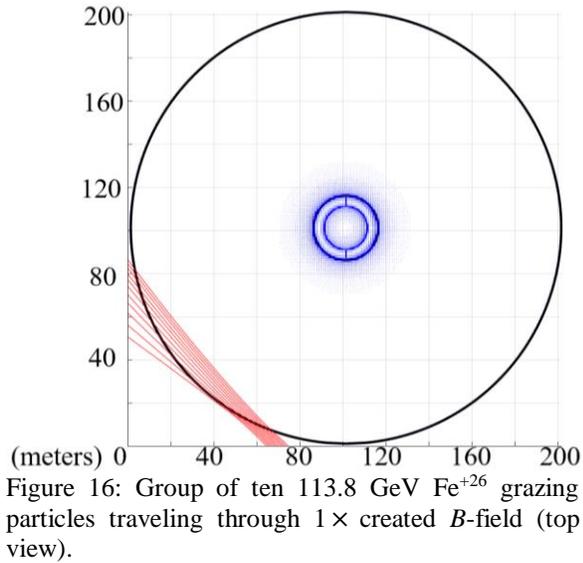

Figure 16: Group of ten 113.8 GeV $Fe^{+26}$ grazing particles traveling through 1× created *B*-field (top view).

The paths of $Fe^{+26}$, $C^{+6}$, $He^{+2}$, $H^+$, and $e^-$ particles for different *B*-field multipliers strength as described earlier were calculated. The results are shown in Tables 3-8.

| *B*-field Multiplier | Mass of Wires (kg) | Length of wire (km) | Critical Current (A) |
|---|---|---|---|
| 0.25 | $1.73 \times 10^4$ | $5.45 \times 10^2$ | 936 |
| 0.5 | $3.63 \times 10^4$ | $1.09 \times 10^3$ | 890 |
| 1 | $8.08 \times 10^4$ | $2.18 \times 10^3$ | 800 |
| 2 | $2.10 \times 10^5$ | $4.36 \times 10^3$ | 617 |
| 4 | $1.02 \times 10^6$ | $8.72 \times 10^3$ | 253 |

Table 3 The integer multiples of the *B*-field with corresponding wire mass and wire length at 16 K

In Table 3, the *B*-field Multiplier is the number multiplied to each vector in the created field. For instance, a *B*-field Multiplier of 1 is no change in the magnitude of the vectors from the initial parameters. A *B*-field Multiplier of 2 is twice the magnitude of each vector in the *B*-field created from the necessary wires.

| *B*-field Multiplier | Max KE $Fe^{+26}$ (GeV) | Synchrotron Radiation (eV) | Plasma Effect Percentage Needed* |
|---|---|---|---|
| 0.25 | 12.0 | $4.99 \times 10^{-8}$ | 85.7% |
| 0.5 | 34.5 | $1.45 \times 10^{-7}$ | 58.9% |
| 1 | 116 | $7.58 \times 10^{-5}$ | none |
| 2 | 229 | $2.80 \times 10^{-4}$ | none |
| 4 | 504 | $2.00 \times 10^{-2}$ | none |

Table 4 Data for $Fe^{+26}$, including the particle's kinetic energy, energy of synchrotron radiation produced, and *the percentage of the test particle's kinetic energy that would require some other method, such as plasma considerations, to deter the particle

In Table 4, "Max KE $Fe^{+26}$" is the maximum grazing particle kinetic energy in GeV that the *B*-field for the given multiplier can deflect. The Synchrotron Radiation is the total produced synchrotron radiation during the particles path.

An accurate idea of how much plasma is induced from the deployable loops is not achievable without real world experimentation. When a particle enters a plasma, its kinetic energy will change. The Plasma Effect Percentage Needed in Table 4 is the percentage decrease of the initial kinetic energy of the particle after it traverses the plasma. So, an 84 GeV $Fe^{+26}$ particle must decrease its kinetic energy to 34.5 GeV in the plasma, which is a 58.9% decrease from 84 GeV, to be fully shielded with a *B*-field Multiplier of 0.5. If no plasma is induced from the deployable loops, a Multiplier of one is required to fully shield the particle. This table gives an idea of the induced plasma's shielding power required to stop the test particle for each *B*-field Multiplier.

$C^{+6}$, $He^{+2}$, $H^+$, and $e^-$ particles were sent into the same *B*-field multipliers to find their maximum kinetic energies and produced synchrotron radiation. The results are in Tables 5-8.

| *B*-field Multiplier | Max KE $C^{+6}$ (GeV) | Synchrotron Radiation (eV) |
|---|---|---|
| 0.25 | 2.56 | $5.00 \times 10^{-10}$ |
| 0.5 | 8.33 | $1.46 \times 10^{-8}$ |
| 1 | 25.1 | $3.90 \times 10^{-6}$ |
| 2 | 53.7 | $2.79 \times 10^{-5}$ |
| 4 | 117 | $1.60 \times 10^{-3}$ |

Table 5 Data for $C^{+6}$

| $B$-field Multiplier | Max KE He$^{+2}$ (GeV) | Synchrotron Radiation (eV) |
|---|---|---|
| 0.25 | 0.981 | 3.99 ×10$^{-10}$ |
| 0.5 | 2.92 | 9.87 ×10$^{-9}$ |
| 1 | 7.90 | 4.32 ×10$^{-7}$ |
| 2 | 17.9 | 3.10 ×10$^{-6}$ |
| 4 | 39.0 | 1.79 ×10$^{-4}$ |

Table 6 Data for He$^{+2}$

| $B$-field Multiplier | Max KE H$^+$ (GeV) | Synchrotron Radiation (eV) |
|---|---|---|
| 0.25 | 0.690 | 3.04 ×10$^{-9}$ |
| 0.5 | 1.96 | 9.75 ×10$^{-8}$ |
| 1 | 4.47 | 7.91 ×10$^{-7}$ |
| 2 | 9.76 | 4.67 ×10$^{-5}$ |
| 4 | 22.1 | 2.35 ×10$^{-2}$ |

Table 7 Data for H$^+$

| $B$-field Multiplier | Max KE e$^-$ (GeV) | Synchrotron Radiation (eV) |
|---|---|---|
| 0.25 | 1000+ | 1000 ×10$^9$ |

Table 8 Data for e$^-$

Electrons lose considerably more energy to synchrotron radiation when traveling at higher kinetic energy through a $B$-field. No electron kinetic energies were found that could make it through the $B$-field. At 1000 GeV, the particle was still shielded from the crew area since energy is lost quicker due to the $\gamma^4$ term. Because of this, even with a $B$-field Multiplier of 0.25, the crew area is a forbidden region for all electrons traveling through the toroidal field. This suggests a second plasma made of electrons may form inside the split toroid which would cause unwanted heating of the wires and a greater power requirement to keep the wires cool. This plasma, however, has a potential to be used as a power source, but this requires further consideration. Some particles may make it through the split in the toroid, but only if they are incident tangential to the toroid at the slit. The orientation of the ship to the Sun will drastically reduce the solar wind particles that make it through the slit.

The effectiveness of this shielding method can be seen when compared to Miller and Zeitlin's data [2]. Miller and Zeitlin show a graph of differential flux vs kinetic energy per nucleon of Fe$^{+26}$, C$^{+6}$, He$^{+2}$, and H$^+$ particles. Estimates from the graphs in Miller and Zeitlin's paper for most probable particle and maximum kinetic energy are provided in Table 9.

| Particle | Most Probable KE | Maximum KE |
|---|---|---|
| Fe$^{+26}$ | 42 GeV | 2800 GeV |
| C$^{+6}$ | 9.0 GeV | 12,000 GeV |
| He$^{+2}$ | 6.0 GeV | 4000 GeV |
| H$^+$ | 0.75 GeV | 5000 GeV |

Table 9 Estimates of the probable kinetic energies and maximum kinetic energies

Using the data from Miller and Zeitlin [2], the most probable kinetic energy GCR particles can be fully deflected using a $B$-field Multiplier of 1. Additionally, Mewaldt [16] notes SEP proton kinetic energy has not exceeded much higher than 1 GeV during the past few major solar flares from 1956-1989. According to Cline and McDonald, the kinetic energy of electrons during solar flares are 3-12 MeV [17]. Even with a $B$-field multiplier of 0.5, the most probable kinetic energies of GCR particle that was simulated can be fully deflected, along with every observed proton and electron from recent solar flares. There will be some particles that cannot be defected, but they have significantly less flux than the average GCR particle and may be deflected by a combination of plasma, $B$-field, and BNNT shielding. The produced synchrotron radiation may be partially absorbed by photovoltaic cells and used as power for onboard equipment [14]. The rest will have to be deflected or absorbed by the BNNT structure housing the crew area [3].

It is important to understand how high energy particles will interact with this design when unshielded. Fig. 17 shows 5 $Fe^{+26}$ with energies increasing to and including 2800 GeV.

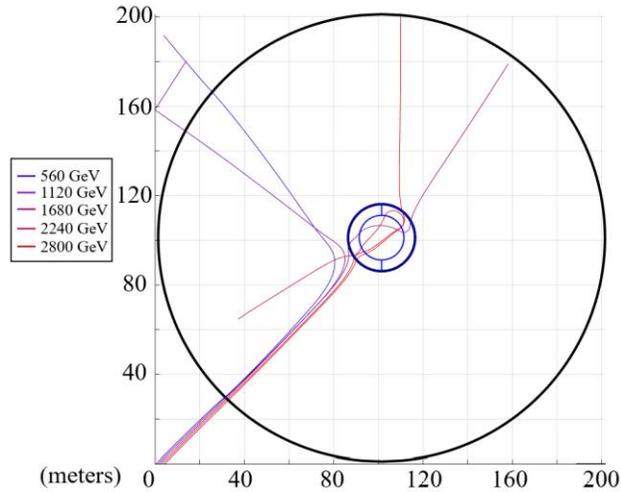

Figure 17: $Fe^{+26}$ particles with increasing energy interacting with a *B*-field Multiplier of 4

Fig. 17 shows that some high energy particles are not shielded. Furthermore, substantial synchrotron radiation will be produced and needs to be considered. These values are in Table 9.

| $Fe^{+26}$ KE (GeV) | Synchrotron Radiation (eV) |
|---|---|
| 560 | $1.82 \times 10^{-4}$ |
| 1120 | $1.26 \times 10^{-2}$ |
| 1680 | $8.59 \times 10^{-1}$ |
| 2240 | 6.12 |
| 2800 | 6.65 |

Table 9: Synchrotron radiation produced from $Fe^{+26}$ particles traveling through a *B*-field Multiplier of 4

With the addition of a plasma surrounding the ship and BNNT around the crew area, these very high energy particles may become shielded from the crew area. Knowing the characteristics of the induced plasma and the shielding effectiveness of BNNTs is essential to finding the limit of this method's shielding ability.

Additionally, a lightweight method to keep the wires under their critical temperature will be required. The addition of well-designed sunshields may reduce the energy required to keep the wires below their critical temperature.

## 5. Conclusion

The simulation shows significant shielding effectiveness of this design. A wire mass of $8.08 \times 10^4$ kg is a reasonable mass considering this is less than the mass of the ISS. Although the mass of the cooling system required to keep the Ti-$MgB_2$ wires under their critical temperature is not included in this calculation, it is important to note the potential of producing thrust from redirected radiation may make this mass acceptable. A space station similar to the ISS could be designed to move between orbits of Earth, the Moon, Mars, or many other bodies using this shielding method without the need for chemical propellants by using the momentum from redirected radiation. The induced electron plasma fields and photovoltaic cells have the potential of powering the systems onboard the ship making it a self-sufficient mobile station. Its cylindrical crew area also has potential to be spun simulating a gravitational field and so reduces problems arising from prolonged exposure to the absence of gravity. This Magnetically Shielded Self-Sufficient Space Station (M5S) holds potential to be a more practical method for interplanetary travel.

These results are promising, but more considerations need to be made. The induced plasma and its added shielding must be studied further. Since calculations of induced plasma are roughly estimated based on knowledge of the plasma trapped in Earth's magnetic field, a real-world experiment may be needed to get an idea of the plasma's characteristics. This could be accomplished using a CubeSat with a deployable wire loop. Considerations must be made of the time required for the plasma to form in the solar wind, the heating of the wires due to the plasma and synchrotron radiation, the produced neutrons from the collision of radiation with any mass on the ship, and the effectiveness of BNNT as a passive shielding method.